\documentclass[aps,prl,groupedaddress,twocolumn,showpacs,showkeys,amssymb]{revtex4}
\usepackage{graphicx}

\begin{document}

\title{Density of states of disordered systems with a finite correlation length}
\author{J. C. Flores$^{a,b}$ and M. Hilke$^{c}$}
\affiliation{$^{a}$ Departamento de F\'{i}sica, Universidad de
Tarapac\'{a}, Casilla 7-D,
Arica, Chile\\
$^{b}$CIHDE, Universidad de Tarapac\'{a}, Casilla 7-D,
Arica, Chile\\
$^{c}$Physics Department, McGill University, 3600 rue University,
Montr\'{e}al (Qu\'{e}bec) H3A 2T8, Canada }

\begin{abstract}
We consider a semiclassical formulation for the density of states
(DOS) of disordered systems in any dimension. We show that this
formulation becomes very accurate when the correlation length of
the disorder potential is large. The disorder potential does not
need to be smooth and is not limited to the perturbative regime,
where the disorder is small. The DOS is expressed in terms of a
convolution of the disorder distribution function and the
non-disordered DOS.  We apply this formalism to evaluate the
broadening of Landau levels and to calculate the specific heat in
disordered systems.

\end{abstract}



\maketitle

Strong disorder plays an important role, particularly in amorphous
systems such as semiconductors or glasses. In electronic systems
and in waves in general, localization due to disorder is common
and has been studied for many years \cite{PWA,Lee,Soukoulis}. It
was shown that localization can strongly affect transport
properties, such as inducing metal-insulator transitions. By
contrast, thermodynamical properties like the specific heat only
depends on the DOS and not on the physics of localization.

Evaluating the DOS for disordered systems is usually very
difficult and is often done with weak disorder expansion
techniques. However, some exact results exist and include the
smoothing out of band edges or singularities \cite{Wegner} and the
existence of Litshitz tails \cite{Lifshitz}. In general, the DOS
decays exponentially close to the band edge for most disorder
distributions \cite{Pastur}. However, there are no general results
valid for the entire DOS in the case of strong disorder and
particularly in dimensions larger than one. Furthermore, most of
the results are based on the assumption of an uncorrelated
disorder potential. Indeed, when a non-zero correlation length
$\xi$ is introduced in the disorder potential the situation is not
well understood. This is nonetheless important for the
understanding of many physical systems, since the correlation
length is often non-zero. For example, in quenched disordered
systems like amorphous metals the nature of the disorder is
largely dependent on the cooling rate and subsequent annealing
procedure \cite{Elliott}. In amorphous semiconductors the
correlation length depends on deposition conditions
\cite{Quicker}, whereas in GaAs/AlGaAs heterostructures the
correlation length can be modified by bias cooling
\cite{Coleridge}.

In this letter we reformulate the semiclassical derivation of the
$DOS$ and show that it is a very good approximation to the exact
result in the limit where the correlation length $\xi$ of the
disorder potential is large. Our approach is based on a
semiclassical formulation, which originated with Kane \cite{Kane}
and was further developed by Ziman \cite{Ziman}. More recently, a
semiclassical approach was used in the context of amorphous
semiconductors \cite{OLeary} and in 1D \cite{Flores03}. Here we
introduce a more general situation, where we consider systems,
where the volume $\tau (E)$ of energy $E$ in the phase-space can
be partitioned for any given realization like:

\begin{equation}
\tau (E)=\sum_{i}\tau _{i}^{f}(E-V_{i}),\label{initial}
\end{equation}
where $\tau _{i}^{f}(E-V_{i})$ is a volume of energy $E-V_{i}$ of
the partition $i$ of the total volume $\tau (E)$. Here $\tau
_{i}^{f}$ corresponds to the volume {\it free of disorder}, since
in this partition $V_{i}$ is assumed approximately constant for a
given realization of disorder.

While the assumption in (\ref{initial}) is quite general and can
be applied to a number of systems, we will focus our attention on
the case of a particle with energy $E$ in a random potential. The
potential is constant by pieces $V_i$ over a constant size $d$,
which represents the correlation length, $\xi$, of the disorder
potential. The Schr\"odinger equation in any dimension is then
simply $
E\psi =H_{f}\psi +\psi\sum_{i}V_{i}(\vec{x}-\vec{a}%
_{i})$, where $H_{f}$ is the Hamiltonian of the {\it disorder
free} system and the functions $V_{i}$ are random independent
functions with a given probability distribution centered at
$\vec{a_i}$. The potential energy $V_{i}$ is approximately
constant for a given realization. In this case, the total volume
$\tau (E)$ can be written as a function of the sub-volume like in
eq.(\ref{initial}).

Semiclassically, the total number of states $N(E)$ is then
proportional to the volume $\tau (E)$ in phase space. Namely,
$N(E)=\tau (E)/h$ \cite{Flores03}. This is our main starting point
to obtain an expression for the DOS of disordered systems. Further
below, we will then analyze under which conditions (\ref{initial})
holds. We assume that $\left\{ V_{i}\right\}$ is an ensemble of
random independent variables with known probability distribution
$P_{tot}=\prod P(V_{i})$. The disorder average $\langle \cdot
\rangle $ of (\ref{initial}) is given by $\langle \tau (E)\rangle
=\sum_{i}\langle \tau _{i}^{f}(E-V_{i})\rangle $ and since the
distribution of the random independent variables $V_{i}$ is the
same, we have $\langle \tau (E)\rangle =\sum_{i}\int dVP(V)\left\{
\tau _{i}^{f}(E-V)\right\}$ and $\langle \tau (E)\rangle =\int
dVP(V)\left\{ \sum_{i}\tau _{i}^{f}(E-V)\right\}$. Because the
summation $\sum_{i}\tau _{i}^{f}(E-V)$ corresponds to the volume
of the disorder-free system, namely,
$\tau^{f}(E-V)=\sum_{i}\tau_{i}^{f}(E-V)$, the expression for the
integrated DOS $N(E)=\tau (E)/h$ becomes
\begin{equation}
\left\langle N(E)\right\rangle =\int dVP(V)N^{f}(E-V)\label{INT}
\end{equation}
and for the DOS $g(E)=\partial N/\partial E$ we have
\begin{equation}
\left\langle g(E)\right\rangle =\int dVP(V)g^{f}(E-V)\label{DOS}.
\end{equation}

\begin{figure}[b]
\includegraphics[scale=0.30]{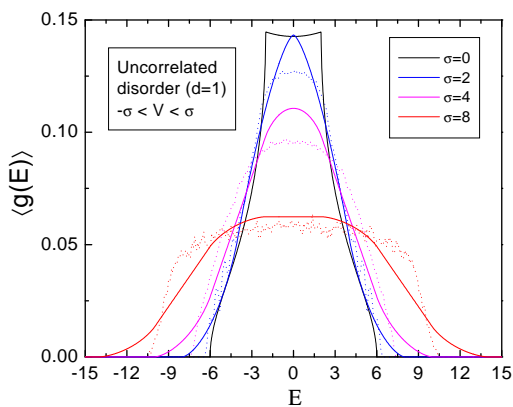}
\caption{Dotted lines: average density of states calculated for a
3D system of size $16^3$ and averaged over 20 configurations for
the Anderson model with a uniform distribution of width
$\sigma=4,8,16$. The bandwidth of the non-disordered case is 12 in
these units. The solid lines are the expressions obtained from
(\ref{DOS}).}

\label{uncorr}
\end{figure}

The disorder-free case corresponds to Dirac's distribution
$P(V)=\delta (V)$, which leads to $\left\langle g(E)\right\rangle
=g^{f}(E)$ as expected. This expression can also be obtained by
using the summation of local density of states following Kane's
original work \cite{Kane}. In what follows we analyze under what
conditions expressions (\ref{INT},\ref{DOS}) can be used and when
they are accurate. We will start by considering the Anderson model
\cite{PWA}, i.e., a discretized version of the Schr\"odinger
equation, which is $\sum_j T_{ij}\psi_j=(E-V_i)\psi_i$, where
$T_{ij}=-1$ for nearest neighbors and 0 otherwise and $V_i$ is our
random potential. The DOS of the non-disordered cubic case is then
simply given by
\begin{equation}
g(E)\sim\int_{(2\pi)^D}\delta(E+2\sum_{i=1}^D
\cos(k_i))d^D\bold{k},\label{DOSTBM}
\end{equation}
where $D$ is the dimension. While evaluating the integral
(\ref{DOSTBM}) gives no simple expression for $g(E)$ for higher
dimensions, $g(E)$ is nonetheless a symmetric function of $E$ and
the DOS near the band edges $E_0=\pm 2D$ scales simply as
$g(E)\sim|E- E_0|^{(D-2)/2}$ like in the continuous case. When
introducing disorder, the $DOS$ broadens.

We first considered the case of uncorrelated disorder, i.e.,
$\langle V_iV_j\rangle\sim\delta_{i,j}$, and where the $V_i$'s are
uniformly distributed between $-\sigma$ and $\sigma$. For this
case (D=3) there exists no analytical result for the entire shape
of the $DOS$. Hence, we evaluated the DOS numerically and show the
result in figure \ref{uncorr} for different strengths of disorder.
In the same figure we also show the DOS evaluated with
(\ref{DOS}). Clearly the agreement is not good. This is not too
surprising and will be discussed below.

\begin{figure}[b]
\includegraphics[scale=0.30]{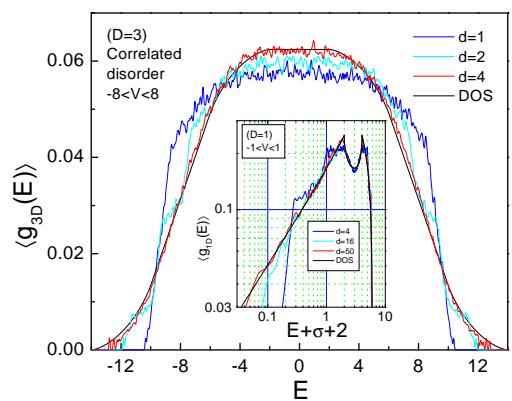}
\caption{Average density of states calculated for a system of size
$16^3$ and averaged over 30 configurations for the Anderson tight
binding model \cite{PWA} with a uniform distribution of width
$\sigma=16$ and $d=1,2,4 $. The black line is the expression
obtained from (\ref{DOS}). In the inset the 1D density of states
is shown for d=4, 16, 50 and the black line is obtained from
(\ref{DOS}).} \label{corr}
\end{figure}

The situation changes quite dramatically, when we include the
condition for the derivation of our semiclassical expression
(\ref{DOS}). Indeed, when we impose that the potential is constant
by pieces, we obtain a much better agreement between expression
(\ref{DOS}) and the exact diagonalization (see figure \ref{corr}).
In figure \ref{corr} we show the 3D DOS for the case where $V_i$
is constant over a fixed interval, ranging from 1 to 4. The larger
the interval the better the agreement. It is important to note
that increasing the interval $d$ does not reduce the strength of
the disorder $\langle V^2\rangle$, but it only increases the
disorder correlation length, where $\xi=d$. While the overall
quality of the fit is quite remarkable as soon as $d$ exceeds 2,
the band tails, however, are only accurate for $|E-E_{min}|>\delta
E$, where $E_{min}$ corresponds to the band edge of expression
(\ref{DOS}). $\delta E$ can be estimated from the cut-off of the
wavelength due to $d$, hence $\delta E\simeq
E(k_{min}+2\pi/d)-E(k_{min})$, where $E(k)$ is the dispersion
relation of the disorder free system. We checked this by
evaluating the 1D case, which is shown in the inset of figure
\ref{corr}, where the accuracy of the band tail DOS is improving
drastically for $d$ increasing from 4 to 50. It is also
interesting to note that the sharp structures close to the band
center are accurately described by (\ref{DOS}) for $d$ large
enough.

\begin{figure}[b]
\includegraphics[scale=0.30]{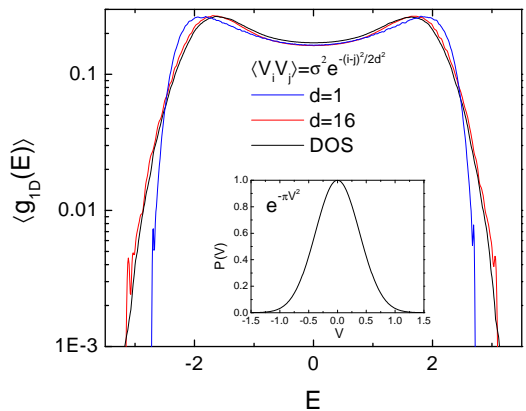}
\caption{Exact average DOS in 1D for a smooth disorder potential
with characteristic length $d=1$ and $d=16$ and the DOS obtained
from (\ref{DOS}). The distribution of the random potential for
$d=16$ is shown in the inset.} \label{gauss}
\end{figure}

The disorder potential considered above, while long range in
nature, is different from the often considered "smooth" disorder
case. For comparison, we considered a smooth potential of the
form: $W_i=(d\sqrt{\pi/2})^{-1/2}\sum_jV_je^{-(i-j)^2/d^2}$, where
$\langle W^2\rangle=\langle V^2\rangle$ and the correlation length
(FWHM) is given by $\xi=2.35 d$. The distribution function for
this potential at large $d$ is simply given by
$P(W)=\sqrt{2\pi\langle V^2\rangle}e^{-W^2/(2\langle V^2
\rangle)}$. We evaluated numerically the exact DOS in 1D and
compared it to expression (\ref{DOS}) for different values of $d$
using the exact numerical distribution function. The results are
shown in figure \ref{gauss}. Here again, expression (\ref{DOS}) is
in excellent agreement with the exact result when $d$ is large.

For smooth disorder this is not too surprising, since the
semiclassical approximation is accurate for a very smooth
potential. Indeed, a recent perturbative calculation in 1D shows
that the semiclassical approximation is exact in the limit, where
$\langle(\nabla V)^2\rangle/\langle V^2\rangle^{3/2}\ll 1$
\cite{Quang}. They also obtained the first order correction of the
DOS to the semiclassical result. In this letter, however, we push
the argument much further and show that for a disordered system
the semiclassical expression (\ref{DOS}) is accurate in the limit
where the correlation length is large, {\it independently} of the
smoothness and {\it dimension} of the system under consideration.
Therefore, we have an expression, which provides a simple form for
the DOS of disordered systems with a large correlation length.
This is the main result of this letter. In localization physics,
disorder correlations can also have a dramatic effect on the
extend of the wave functions and even lead to the existence of
extended states in 1D \cite{flores} and 2D \cite{hilke}. Without
correlations, Lloyd showed that for $P(V)$ a Lorentz distribution,
the exact DOS is also given by expression (\ref{DOS})
\cite{Lloyd}.

We now turn to some important consequences of expression
(\ref{DOS}). We start with the singularities in the disorder-free
system, which will be smoothed out if the disorder distribution is
continuous. An interesting example of this is the situation of a
perpendicular magnetic field in two dimensions, where the spectrum
is discrete and the density of states is given by a sum of
$\delta$ functions centered at the Landau levels, i.e., $g(E)\sim
\sum_n\delta(E-(n+1/2)\hbar \omega_c)$, where $\omega_c$ is the
cyclotron frequency. Using expression (\ref{DOS}) to obtain the
average DOS in the presence of disorder immediately gives a
broadening of the Landau levels only related to the disorder
strength and the disorder distribution. Hence,
\begin{equation}
\langle g(E) \rangle\sim \sum_n P(E-(n+1/2)\hbar\omega_c)
\end{equation}
It is interesting to note that this Landau level broadening does
not depend on the Landau level index. While the exact shape of the
Landau level broadening is still under debate, experiments on
GaAs/AlGaAs heterostructures, which have long range disorder,
indicate that the broadening is magnetic field dependent at low
fields, where the Landau level overlap is significant. At higher
fields the broadening becomes independent of the field and of the
Landau level index and the shape is found to be best fitted by a
Lorentzian \cite{Potts}. Hence, expression (\ref{DOS}) is
consistent with experimental findings for large fields in systems
with long range disorder.

Another interesting consequence of expression (\ref{DOS}) is the
widening of the bands. Indeed, if $P(V)$ has a finite support,
namely, $-\sigma <V<+\sigma $, then the bands will be extended by
$\sigma$ on each side. This can lead to the decrease of gaps in
disordered materials. The band edges will acquire tails.
Expression (\ref{DOS}) leads to tails, which depend on the
disorder distribution. For continuous systems the disorder free
band edge is given by $g(E)\sim \sqrt{E}$ in $D=3$. Introducing
disorder will modify this dependence. It is quite straightforward
to see that, for a square distribution $P(V)=1/2\sigma$ for
$-\sigma <V<+\sigma$ and $P(V)=0$ otherwise, we obtain in any
dimension
\begin{equation}
\left\langle g(E)\right\rangle =\frac{1}{2\sigma }\left\{
N^{f}(E+\sigma )-N^{f}(E-\sigma )\right\} ,\label{gsquare}
\end{equation}
where $N^{f}$ is the integrated DOS of the disorder-free system.
Therefore, $\langle g(E)\rangle\sim (E+\sigma)^{3/2}$ at the band
edge and the band tail is not exponential for this distribution.
Numerically, we see that this is indeed the case for the
correlated case ($d=50$) shown in the inset of figure \ref{corr}.
For other disorder distributions, $P(V)$, such as Gaussian or
Poisson, we do recover an exponential band tail. Fundamentally,
the disorder distribution will govern the dependence of the band
tails. In the high disorder limit $E/\sigma \ll 1$, $ \left\langle
g(E)\right\rangle \simeq \frac{1}{2\sigma }N^{f}(\sigma )$ is
independent on energy.

We now turn to the specific heat of disordered systems. The
internal energy $U_{F}$ and the particle number $N_{F}$ for
fermions is given by $ U_{F} = \int \frac{E}{e^{(E-\mu
)/KT}+1}\text{ }\langle g(E)\rangle dE$ and $ N_{F} = \int
\frac{1}{e^{(E-\mu )/KT}+1}\text{ }\left\langle g(E)\right\rangle
dE$. As above we consider the disorder-free DOS $g(E)=\alpha
\frac{D}{2}E^{\frac{D-2}{2}}$, where $\alpha $ is a constant. In
the high temperature limit (Boltzmann distribution) and using
(\ref{gsquare}) leads to a specific heat
$C_{Boltz}=\frac{\partial }{\partial T}$ $\left( \frac{%
U_{Boltz}}{N_{Boltz}}\right) $ or
\begin{equation}
C_{Boltz}=K\left\{ \frac{D+2}{2}-\left( \frac{\sigma }{KT\sinh \sigma /KT}%
\right) ^{2}\right\} .
\end{equation}
For the case $D=1$ we recover the results found in reference
\cite{Flores03}. For all dimensions, the limit $\sigma \rightarrow
0$ gives $C_{Boltz}=KD/2$ in agreement with the Dunlap-Petit law.
The case $\sigma \rightarrow \infty $ leads to
$C_{Boltz}=K(D+2)/2$, which represents an enhancement of the
specific heat due to disorder.

\begin{figure}[b]
\includegraphics[scale=0.23]{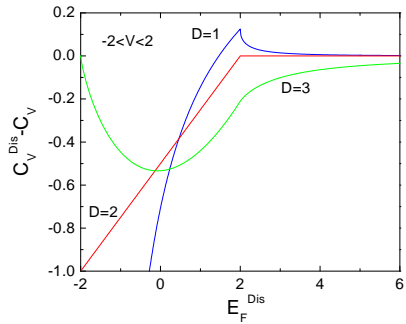}
\caption{Difference between the specific heat of the disordered
system and the ordered system in units of $\frac{(\pi
K)^2T}{3}\alpha$ as a function of the Fermi level of the
disordered system.} \label{CVfig}
\end{figure}

In the opposite limit of low temperatures, where we can use the
Sommerfeld expansion, the difference between the specific heat of
the disordered system and the ordered system (for the same
electronic density) is given by

\begin{equation}
C_D-C_V\simeq\frac{(\pi K)^2T}{3}\frac{\alpha
E_D^{\frac{D-6}{2}}\sigma^2}{12}(2-D)D,
\end{equation}
where $E_D$ is the Fermi level of the disordered system and we
further assumed that $\sigma\ll E_D$. Clearly this expression
shows that the disorder decreases the specific heat in 3D but
enhances it in 1D. The 3D case is simple to understand, since at
low temperatures the specific heat is mainly determined by the
tail of the DOS. Hence, for a given energy bandwidth, the DOS of
the disordered case is smaller than the DOS of the ordered case,
which leads to a smaller specific heat in the disordered case. The
1D result is more surprising, but can be understood by looking at
all relative disorder strengths presented in figure \ref{CVfig}.
Indeed, at high disorder strength, (i.e. $E_D$ small), where the
low energy DOS is dominated by the tails, we do recover that the
specific heat is smaller in the disordered case than in the
ordered case. In the opposite limit of small relative disorder the
situation is reversed because the disorder-free DOS in 1D
decreases with energy, leading to a comparatively larger specific
heat for the disordered case.

In conclusion, we have shown that the average DOS of disordered
systems can be derived in a semiclassical way to represent an
accurate DOS for a disorder potential when the correlation length
is large. We further, evaluated a few physical quantities such as
the broadening of the Landau levels, the characterization of band
edges and tails and finally, discussed the electronic specific
heat in these systems.

M.H. would like to acknowledge support from RQMP, FQRNT and NSERC.

\bibliography{refs}

\begin{thebibliography}{99}
\bibitem{PWA}  P.W. Anderson, Phys. Rev. {\bf 109}, 1492 (1958).

\bibitem{Lee}  P.A. Lee and T.V. Ramakrishnan, Rev. Mod. Phys. {\bf 57}, 287
(1985).

\bibitem{Soukoulis}  C.M. Soukoulis and E.N. Economou, Wave Random Media {\bf 9},
255 (1999).

\bibitem{Wegner} F.J. Wegner, Phys. Rev. B {\bf 19}, 783 (1979).

\bibitem{Lifshitz} I.M. Lifshitz, Sov. Phys. Usp. {\bf 7}, 549 (1965).

\bibitem{Pastur} L. Pastur and A Figotin, Spectra of Random and
Almost periodic operators, Springer Verlag, (1992).

\bibitem{Elliott} S.R. Elliott, Physics of amorphous materials,
John Wiley \& Sons (1989).

\bibitem{Quicker} D. Quicker and J. Kakalios, Phys. Rev. B {\bf
60}, 2449 (1999).

\bibitem{Coleridge} P.T. Coleridge, Semicond. Sci. Technol. {\bf
12}, 22 (1997).

\bibitem{Kane} E. O. Kane, Phys. Rev. {\bf 131}, 79 (1963).

\bibitem{Ziman}  J. M. Ziman, {\it Models of Disorder} (Cambridge University
Press, 1979).


\bibitem{OLeary} S.K. O'Leary, S. Zukotynski and J.M. Perz,
Phys. Rev. B {\bf 51}, 4143 (1995).

\bibitem{Flores03}  J. C. Flores, Phys. Stat. Sol. (b) {\bf 239}, 168 (2003).

\bibitem{Quang} D.N. Quang and N.H. Tung, Phys. Rev. B {\bf 60},
(1999) 13648.

\bibitem{flores}J.C. Flores, {\em J. Phys. Condens. Matter} {\bf 1}, 8471
(1989) and D.H. Dunlap, H.-L. Wu, and P.W. Phillips, {\em Phys.
Rev. Lett.} {\bf 65}, 88 (1990).

\bibitem{hilke} M. Hilke, {\em J. Phys. A: Math. Gen.} {\bf 27}, 4773
(1994) and M. Hilke, {\em Phys. Rev. Lett.} {\bf 91}, 226403
(2003).

\bibitem{Lloyd} P. Lloyd, J. of Phys. C {\bf 2}, 1717 (1969).

\bibitem{Potts} A. Potts {\em et al.}, J. of Phys.: Condens. Matter {\bf 8}, 5189(1996).

\end{thebibliography}

\end{document}